\documentclass[12pt,epsf]{article}
\usepackage{graphicx}
\usepackage{amsmath}
\usepackage{amsfonts}
\begin{document}

\def\a{\alpha}
\def\b{\beta}
\def\c{\varepsilon}
\def\d{\delta}
\def\e{\epsilon}
\def\f{\phi}
\def\g{\gamma}
\def\h{\theta}
\def\k{\kappa}
\def\l{\lambda}
\def\m{\mu}
\def\n{\nu}
\def\p{\psi}
\def\q{\partial}
\def\r{\rho}
\def\s{\sigma}
\def\t{\tau}
\def\u{\upsilon}
\def\v{\varphi}
\def\w{\omega}
\def\x{\xi}
\def\y{\eta}
\def\z{\zeta}
\def\D{\Delta}
\def\G{\Gamma}
\def\H{\Theta}
\def\L{\Lambda}
\def\F{\Phi}
\def\P{\Psi}
\def\S{\Sigma}

\def\o{\over}
\def\beq{\begin{eqnarray}}
\def\eeq{\end{eqnarray}}
\newcommand{\gsim}{ \mathop{}_{\textstyle \sim}^{\textstyle >} }
\newcommand{\lsim}{ \mathop{}_{\textstyle \sim}^{\textstyle <} }
\newcommand{\vev}[1]{ \left\langle {#1} \right\rangle }
\newcommand{\bra}[1]{ \langle {#1} | }
\newcommand{\ket}[1]{ | {#1} \rangle }
\newcommand{\EV}{ {\rm eV} }
\newcommand{\KEV}{ {\rm keV} }
\newcommand{\MEV}{ {\rm MeV} }
\newcommand{\GEV}{ {\rm GeV} }
\newcommand{\TEV}{ {\rm TeV} }
\def\diag{\mathop{\rm diag}\nolimits}
\def\Spin{\mathop{\rm Spin}}
\def\SO{\mathop{\rm SO}}
\def\O{\mathop{\rm O}}
\def\SU{\mathop{\rm SU}}
\def\U{\mathop{\rm U}}
\def\Sp{\mathop{\rm Sp}}
\def\SL{\mathop{\rm SL}}
\def\tr{\mathop{\rm tr}}
\def\mpl{M_{PL}}

\def\IJMP{Int.~J.~Mod.~Phys. }
\def\MPL{Mod.~Phys.~Lett. }
\def\NP{Nucl.~Phys. }
\def\PL{Phys.~Lett. }
\def\PR{Phys.~Rev. }
\def\PRL{Phys.~Rev.~Lett. }
\def\PTP{Prog.~Theor.~Phys. }
\def\ZP{Z.~Phys. }


\baselineskip 0.7cm

\begin{titlepage}

\begin{flushright}
IPMU 13-0013\\
\end{flushright}

\vskip 1.35cm
\begin{center}
{\bf A Simple Solution to the Polonyi Problem\\ in Gravity Mediation

}

\vskip 1.2cm
Keisuke Harigaya$^1$, Masahiro Ibe$^{2,1}$, Kai Schmitz$^{1}$ and Tsutomu T. Yanagida$^1$
\vskip 0.4cm
$^1${\it Kavli IPMU, TODIAS, University of Tokyo, Kashiwa 277-8583, Japan}\\
$^2${\it ICRR, University of Tokyo, Kashiwa 277-8582, Japan}
\vskip 1.5cm

\abstract{
The Polonyi field is a necessary ingredient in any viable scenario of
gravity mediated supersymmetry breaking.
However, it is known that the presence of the Polonyi field leads to
several serious cosmological problems, which are collectively referred to as the Polonyi problem.
We show that the Polonyi problem can be solved if the Polonyi field
 couples to a pseudo modulus in the superpotential and this pseudo
 modulus has a large field expectation value during inflation.
To illustrate our idea, we construct an explicit model which can be readily
connected to scenarios of gravity mediation.
The generation of the mass parameters contained in our model by strong gauge
dynamics is also commented on.}
\end{center}
\end{titlepage}

\setcounter{page}{2}

\section{Introduction}

The Polonyi problem is one of the most serious problems in the gravity
mediated supersymmetry (SUSY) breaking scenario~\cite{Coughlan:1983ci}. The presence of the
Polonyi field $S$ is an inevitable ingredient in gravity mediation, since otherwise we would
have vanishing gaugino masses at the tree level and
hence gauginos would be much lighter than sfermions. The
non-discovery of the gluino at the LHC~\cite{:2012mfa}
already suggests too large squark masses
assuming gravity mediated SUSY breaking without the Polonyi field (see for example~\cite{:2004di}),
which makes the SUSY solution to the hierarchy problem very questionable.

However, if the Polonyi field $S$ exists we encounter several serious cosmological
problems. This is because the Polonyi field is neutral under any symmetry and
its origin has no meaning, and hence it naturally has an $O(\mpl)$ initial
value at the end of inflation ($\mpl \simeq 2.4\times 10^{18}$ GeV is the
reduced Planck scale). The Polonyi field starts to coherently
oscillate when the expansion rate of the universe becomes smaller than
the Polonyi mass and it decays at much later times. In particular, its
decay produces too much entropy inducing a huge dilution of the
primordial baryon-number asymmetry \cite{Coughlan:1983ci}. Furthermore,
its decay occurs during/after big bang nucleosynthesis (BBN) and
destroys the light elements created during BBN (for a recent analysis,
see \cite{Nakayama:2011zy}).

In this paper, we show that the Polonyi problem can be easily solved if
there is a pseudo-flat direction which couples to the Polonyi field in
the superpotential. A large vacuum expectation value of this flat direction
gives a large mass to the Polonyi field such that it sits close to
the minimum of the potential during inflation.
We show that this minimum during inflation coincides with the minimum of the
potential in the present universe and that the coherent oscillation of the Polonyi
field is thus suppressed.

This paper is organized as follows.
In section \ref{sec:Polonyi problem}, we briefly review the Polonyi
problem.
In section \ref{sec:model}, we present our solution to the Polonyi problem.
The last section is devoted to  discussion and conclusions.
Finally, the generation of the mass parameters of our model by strong gauge dynamics
is discussed in the Appendix.

\section{The Polonyi problem}
\label{sec:Polonyi problem}
In this section we briefly review the Polonyi problem.
The Polonyi field, which is completely neutral under any symmetry, is a
necessary ingredient in any viable scenario of gravity mediated SUSY breaking.
The Polonyi field has a mass $m_S\sim \mu^2/\mpl$, where $\mu$ is the
SUSY breaking scale. $m_S$ is of
the order of the gravitino mass $m_{3/2}={\cal O}({\rm TeV})$.
Since the mass of the Polonyi field is small and negligible in comparison with
the typical Hubble scale of inflation, the
Polonyi field generally has a non-zero vacuum expectation value of
${\cal O}(\mpl)$ during inflation. Here, we define the minimum of the
Polonyi potential in the present universe as $S=0$.

After the end of inflation, when the
Hubble scale becomes smaller than $m_S$, the Polonyi field starts to coherently
oscillate around the minimum of its potential, i.e.\ $S = 0$. We suppose that the
oscillation starts before the reheating process completes, since
otherwise the cosmological problem mentioned below becomes more
disastrous. At the beginning of the oscillation, the ratio of the energy density of the
Polonyi field $\rho_{S}$ to that of inflaton field $\rho_{\phi}$ is given by
\begin{eqnarray}
 \frac{\rho_{S}}{\rho_{\phi}}=\frac{m_S^2
  |S_i|^2}{3\mpl^2m_S^2}=\frac{|S_i|^2}{3\mpl^2}.
  \label{eq:rhoratio}
\end{eqnarray}
Here, we have used the equality $H\simeq m_S$ where $H$ is the Hubble
parameter of the universe.
The ratio in Eq.~(\ref{eq:rhoratio}) is conserved
until the beginning of the reheating process.

After the reheating
process completes, the ratio of the energy density of the Polonyi field
to the entropy density of the universe $s$ is given by
\begin{eqnarray}
 \frac{\rho_S}{s} = \frac{\rho_S}{\rho_\phi}\frac{\rho_\phi}{s}=\frac{|S_i|^2T_{R}}{4\mpl^2}
= 3\times 10^{8}~{\rm GeV}\,\frac{T_R}{10^9~{\rm GeV}}\left(\frac{|S_i|}{\mpl}\right)^2,
\end{eqnarray}
where $T_R$ is the reheating temperature. Since the Polonyi field
couples to the visible sector through dimension-5 operators, its decay
rate can be roughly estimated as
\begin{eqnarray}
\Gamma_S\simeq \frac{m_S^3}{8\pi \mpl^2} = 10^{-5} {\rm sec}^{-1}\,\left(\frac{m_S}{{\rm TeV}}\right)^3.
\end{eqnarray}
Therefore, the Polonyi field decays during or after BBN and its decay
spoils the success of BBN. Thus, the
energy density of the Polonyi field is strictly constrained. For
example, for $m_S=1$ TeV, successful BBN requires $\rho_S/s <
10^{-13}$ GeV \cite{Kawasaki:2004qu}. This constraint leads to
\begin{eqnarray}
 |S_i|< 10^{-11}\mpl\times\left(\frac{T_R}{10^9~{\rm GeV}}\right)^{-1/2},
\label{eq:condition on Si}
\end{eqnarray}
which requires substantial fine-tuning.

This fine tuning is considered to be a fine
tuning of the K\"ahler potential.
Terms in the K\"{a}hler potential such
as $K \supset c'\phi^{\dag}\phi S/\mpl + c \mpl S$, where $\phi$ is the chiral
multiplet of the inflaton and $c'$ and $c$ are coupling constants, are not protected by any symmetry and
generally non-zero. That is, the smallness of these terms is not natural
even in the sense of 't Hooft \cite{'tHooft:1979bh}.
These terms bring about an $\mathcal{O}(\mpl)$ difference between the minimum of the Polonyi
potential during and after inflation because of their coupling to the
potential energy of the inflaton.
The situation is most severe in the case of the term linear in
the Polonyi field $K \supset c\mpl S$. This term is
always generated by the interaction with the gauge multiplets in the model
and it is quadratically divergent~\cite{Ibe:2006am}. Therefore, even if we suppose that the linear
term is tuned to be zero at some energy scale of interest, a small change
of the coupling constants in the theory leads to large linear term. This
behavior should be compared with the Higgs mass term in the standard
model.
Without fine tuning of the linear term in the K\"ahler potential, we are thus never able
to satisfy the condition in Eq.~(\ref{eq:condition on Si}).

\section{A solution to the Polonyi problem}
\label{sec:model}
Let us present a simple model to demonstrate our mechanism to solve the
Polonyi problem. First of all, we assume an $R$ symmetry
throughout this paper to understand the smallness of the constant term
($i.e.$ gravitino mass) in the superpotential, otherwise we would expect a
large gravitino mass of order $\mpl$. We suppose that a condensation of
some field operator ${\cal O}$ breaks the $R$ symmetry and generates the
small constant term in the superpotential, $W_0=\vev{{\cal O}}$.%
\footnote{\label{fn:discreteR}
As discussed in Ref.\,\cite{Dine:2009sw}, the generation of the constant term in the superpotential 
by the breaking of a genuinely continuous $R$-symmetry requires an $R$-charged field having
a vacuum expectation value of the order of the Planck scale and a non-vanishing $F$-term.
In the following, however, we assume instead that, in the sector responsible for the generation
of the constant term $W_0$, the $R$-symmetry is a discrete symmetry or at least
broken by the anomalies of some strong gauge dynamics so that it effectively ends up
being a discrete symmetry.\smallskip}

Now, let us introduce the Polonyi field $S$. Since it has to be completely neutral under
any symmetry in order to generate the gaugino mass terms, it generally couples
to the operator ${\cal O}$ in the superpotential. Thus, we have a superpotential,
\begin{equation}
W= k_1S\vev{{\cal O}} + \vev{{\cal O}} ,
\label{eq:cosmological constant}
\end{equation}
where $k_1$ is a constant and ${\cal O}$ as well as the Polonyi field $S$
carry $R$ charges 2 and 0, respectively.
Here and hereafter, we work in units such that $M_{PL}=1$.
We easily find that a vanishing cosmological constant is realized by
choosing an appropriate value for $k_1$.
This is a big merit of this type of model, in which we
assume only one dynamical origin for the $R$ symmetry and SUSY breaking scales,
namely the condensation of the operator~$\cal O$.
This condensation can be achieved 
by the condensation of hidden gauginos~\cite{Veneziano:1982ah} or quarks such that
${\cal O}=W^\alpha W_\alpha$ or $~(QQ)^n$,
for example.%
\footnote{Since these operators carry $R$ charge, the condensation
induces a spontaneous breakdown of the $R$ symmetry. Since the $R$ symmetry
is explicitly broken down to a discrete one by the gauge anomaly from the outset,
the condensation produces too many domain walls in the early universe. However, as
pointed out in Ref.~\cite{Dine:2010eb}, the domain walls are easily inflated away.
}
The dynamical generation of the
quark condensate
by hidden strong gauge interactions is reviewed in Appendix \ref{sec:IYIT}.
As for the following discussion, we merely assume that the operator $\cal O$
has a non-vanishing expectation value and focus on explaining our main idea.

Notice that the superpotential in Eq.~\eqref{eq:cosmological constant} yields
a completely flat potential in $S$ up to supergravity corrections, as long
as the K\"ahler potential for $S$ is the minimal one, $K = S^\dagger S$.
To resolve this vacuum degeneracy, we introduce pseudo-flat directions in the scalar
potential by adding a pair of chiral multiplets $X$
and ${\bar X}$ which couple to $S$ in the superpotential. However, we
must also introduce an additional pair of fields, $Y$ and ${\bar Y}$,
as mass partners of ${\bar X}$ and $X$, respectively,
in order to stabilize the SUSY breaking vacuum. Now the superpotential is given by
\begin{equation}
W= \mu^2 S + k_2 SX{\bar X} + mX{\bar Y} + m'{\bar X} Y + \langle {\cal
 O'}\rangle,
\label{eq:superpotential}
\end{equation}
where $\mu^2 =k_1\langle {\cal O}\rangle$ and $k_2$ are constants. We
have absorbed a possible mass term $M X{\bar X}$ in the definition of
$S$. The constant term $\vev{\cal O}$ is shifted to $\vev{\cal{O}'}$ in
accordance with this redefinition. 
$m$ and $m'$ are dimension-1 parameters and their origin can also be
obtained from strong gauge dynamics as we explain in Appendix A.
The magnitude of the mass parameters $m$ and $m'$ is discussed in the
following along with our mechanism to solve the Polonyi problem.

Let us discuss the symmetries of the theory.
We take the $R$ charges $X[-1], {\bar X}[3], Y[-1]$, ${\bar Y}[3],
m[0]$ and $m'[0]$. The reason why we choose the $R$ charge of $X$
to be $-1$ is explained later. Furthermore, we introduce another
discrete symmetry $Z_M$ different from the $R$ symmetry
to suppress mass terms such as $W=M'Y{\bar Y}$.
If such a mass term existed, we would expect that $M'\sim 1$.
In this case, $X$ would become
as light as $m m'/M'$ by the seesaw mechanism.
Later, we will assume that $X$ has a large field value during inflation.
Therefore, we would have again a modulus problem if $X$ was such light.
We assume the charges under this discrete $Z_M$ to be $Y[1], {\bar Y}[1]$,
$m[-1], m'[-1]$ and $0$ for all other fields and mass parameters, i.e.\ $S,
X, {\bar X}$ and $\mu$.
We also introduce a matter parity $Z_2$
to guarantee the stability of the lightest supersymmetric particle
so that it can form dark matter as well as to explain the observed long
lifetime of the proton. Under this $Z_2$, quarks, leptons, $X, {\bar
X}, Y$ and ${\bar Y}$ have odd parity and all other fields have even parity.  All
charge assignments are summarized in Table \ref{table:charge assignment}.
We see that the above mentioned dangerous term $W=M'Y\bar{Y}$ is forbidden.

\begin{table}
\begin{center}
 \begin{tabular}{|c||c|c|c|c|c|c|c|c|}
\hline
  & $S$ & $X$ & $\bar{X}$ & $Y$ & $\bar{Y}$& $\mu^2$&$m$& $m'$ \\
\hline
$R$ &$0$&$-1$&$3$&$-1$&$3$&$2$&$0$&$0$\\
$Z_M$ &$0$&$0$&$0$&$1$&$1$&$0$&$-1$&$-1$\\
$Z_2$ &$+$&$-$&$-$&$-$&$-$&$+$&$+$&$+$\\
\hline
 \end{tabular}
\caption{Charge assignments for the fields $S$, $X$, $\bar{X}$, $Y$, and $\bar{Y}$
as well as the spurious fields $\mu^2$, $m$, and $m'$ under the $R$ symmetry,
the discrete symmetry $Z_M$ and the matter parity $Z_2$.}
\label{table:charge assignment}
\end{center}
\end{table}

For the time being, we neglect possible higher-dimensional terms in the
superpotential. In the SUSY breaking vacuum,
\begin{equation}
\langle X\rangle = \langle {\bar X}\rangle=\langle Y\rangle=\langle {\bar Y}\rangle =0,
\label{eq:XXYY}
\end{equation}
and $S$ is undetermined at the tree level.
Here, we assume that $|k_2\mu^2|<|m^2|,|m'^2|$ since otherwise the SUSY breaking
vacuum would correspond to $\vev{Y}=\vev{\bar{Y}}=0,$ $\vev{X}=\vev{\bar{X}}\sim
\mu/k_2^{1/2}$ along with a vanishing $F$-term of $S$.
The flat direction $S$ is
lifted by one-loop
corrections due to its interaction with $X$ and ${\bar X}$.
In fact, by integrating out the massive fields $X,~\bar{X},~Y$, and
$\bar{Y}$, we obtain the following correction to the K\"ahler potential,
\begin{eqnarray}
 \delta K = -\frac{k_2^4}{96\pi^2} \left(\frac{S^{\dag}S}{m}\right)^2 + ... \,,
\end{eqnarray}
which generates a positive mass squared for $S$ around $S=0$.
Here, we take $m=m'$ for simplicity.
Thus, we have a unique SUSY breaking
vacuum, in which
\begin{equation}
\langle S\rangle=0,
~~~F_S=\mu^2,
\end{equation}
along with Eq.~\eqref{eq:XXYY}.

Now we are at the point to show our solution of the Polonyi problem.
Suppose that the Hubble constant during inflation $H_{\rm inf}$
is larger than the mass scales $m$ and $m'$.%
\footnote{Therefore, it is required
that
$H_{\rm inf} > |m|,|m'| > |\mu| = \left(\sqrt{3}m_{3/2}\right)^{1/2}
= 6 \times 10^{10}~{\rm GeV} \left(\frac{m_{3/2}}{1~{\rm TeV}}\right)^{1/2}$.
}
We also assume that only $X$ obtains a Hubble-induced negative mass squared
and hence rolls down to the Planck scale $\langle X\rangle \sim 1$
during inflation. Then, $S$ obtains a large SUSY-invariant mass of the
order of the Planck scale and is forced to sit around the origin $S=0$.
Due to the term linear in $S$ in the K\"{a}hler potential, the field value is
not exactly one but $\langle S\rangle \simeq H_{\rm inf}^2/\langle
X\rangle\sim H_{\rm inf}^2$.
After the end of inflation, $X$ rolls down towards the minimum of the
potential forced by the mass $m$. $S$ also rolls down towards the origin $S=0$.
Notice that the potential minimum during inflation, $S\simeq H^2_{\rm inf}$
is close to the true minimum $S=0$ in the present universe and the
Polonyi problem hence does not exist.

The next question is whether it is possible for the field $X$ to
decay without causing any problems.
The decay of $X$ occurs through the following superpotential,
\begin{equation}
W=c_1X{\bar u}{\bar d}{\bar d} + c_2XLL{\bar e},
\end{equation}
where $c_1$ and $c_2$ are coupling constants. $\bar{u},~\bar{d},~L$ and
$\bar{e}$ are the right-handed up-type quark, down-type quark, left-handed lepton
doublet and right-handed charged lepton, respectively.  
Note that we can assign $R$ charge $+1$ to the fields $\bar{u},~\bar{d},~L$ and
$\bar{e}$ as usual without running into a contradiction with the
$SU(5)$ GUT~\cite{Georgi:1974sy} or the
Giudice-Masiero mechanism \cite{Giudice:1988yz}.
It is now clear that the above interactions are consistent with all
symmetries that we have imposed. This is the reason why we previously assigned the
$R$ charge $-1$ to $X$.
The decay rate of the pseudo modulus $X$ is estimated as
\begin{equation}
\Gamma_X \simeq \left(|c_1|^2+|c_2|^2\right)m^3 \sim \left(\frac{m}{m_{S}}\right)^3 \Gamma_S
 \simeq 10^{19} {\rm sec}^{-1} \,\left(\frac{m}{10^{11}~{\rm GeV}}\right)^3,
\end{equation}
which is sufficiently large so that the oscillating field $X$ decays well
before BBN.
\bigskip

Before closing this section, let us comment on the effect of higher-dimensional operators.
Operators including $X$ might spoil our mechanism since $X$ has a large
field value during inflation and higher-dimensional operators containing $X$
are not suppressed by the Planck scale.
Such terms are, however, absent in our model
since we assume a continuous $R$ symmetry in the sector consisting of
the fields $X,~\bar{X},~Y$ and $\bar{Y}$.
If the $R$ symmetry in this sector is a discrete one even in the classical
limit,\footnote{Recall that in the $R$ symmetry-breaking sector
any continuous $R$ symmetry is always explicitly broken down to a discrete
one, if not at the tree level then at least at the quantum level
due to gauge anomalies, see also footnote~\ref{fn:discreteR}.} the interaction
$W \supset X^n\bar{X}$ would exist for an appropriate integer $n$.
This interaction would then force $S$ to roll
down the scalar potential to minimize the $F$-term of $\bar{X}$, which would result in
$\vev{S}\sim 1$ during inflation and restore the Polonyi problem.
However, even if we only assume the discrete $R$ symmetry, we can
invoke the anomaly-free $U(1)_X$ symmetry, which is a linear combination
of $U(1)_Y$ and $U(1)_{B-L}$. Imposing this symmetry is natural rather
than artificial if we believe that $U(1)_{B-L}$ is a gauge
symmetry.

To see how well we can suppress problematic higher-dimensional
operators, let us discuss an $SU(5)$ GUT model including an additional $U(1)_X$
gauge symmetry as well as right-handed neutrinos $N$. The charge assignments
for all fields in this model are shown in Table~\ref{table:charge assignment2}.
$\Phi$ and $\bar{\Phi}$ are the $B$$-$$L$ breaking fields.
When the $R$ symmetry is the $Z_{nR}$,
higher-dimensional superpotentials which are consistent with all the symmetries and
which lead to the above mentioned superpotential are
of the form $(\bar{\Phi}^3X^{10})^\ell X\bar{X}$.
Here, $\ell$ is the positive
integer which satisfies that $10\,\ell$ is a multiple of $n$.
The effects of the higher-dimensional operators are then suppressed by
$(\vev{\bar{\Phi}}/\mpl)^{3l}$. By choosing $n$ appropriately, the Polonyi
field $S$ is well trapped near the origin $S=0$ during and after inflation.

\begin{table}
\begin{center}
 \begin{tabular}{|c||c|c|c|c|c|c|c|c|c|c|c|c|}
\hline
  & $10$ & $\bar{5}$ & $5_H$ & $\bar{5}_H$ & $N$& $\Phi$ &$\bar{\Phi}$& $S$ & $X$ &
  $\bar{X}$ & $Y$ & $\bar{Y}$\\
\hline
$U(1)_X$ &$-1$&$3$&$2$&$-2$&$-5$&$10$&$-10$&$0$&$3$&$-3$&$3$&$-3$\\
$R$&$1$&$1$&$0$&$0$&$1$&$0$&$0$&$0$&$-1$&$3$&$-1$&$3$\\
\hline
 \end{tabular}
\caption{$U(1)_X$ and $R$ charge assignments for all fields in the $SU(5) \times U(1)_X$ model
discussed at the end of section~\ref{sec:model}.}
\label{table:charge assignment2}
\end{center}
\end{table}

\section{Discussion and conclusion}

In this letter, we have proposed a simple SUSY-breaking model in which
the SUSY-breaking vacuum at the origin of the Polonyi field $S$ is the unique
true vacuum, at least as long as Planck-suppressed interactions are neglected.
It is crucial for this model to contain a pseudo-modulus of mass $m$.
We have found that the simple SUSY-breaking model naturally provides a solution
to the Polonyi problem if the mass parameter $m$ is smaller than the
Hubble rate $H_{\textrm{inf}}$ during inflation.
The pseudo-modulus may then have a large field value during inflation
such that the Polonyi field acquires a large
mass and is stabilized near the true minimum of the potential during inflation
as well as in the present universe.
Though we have explained this surprising mechanism by means of a concrete example,
we believe that it is general and also applicable to other models.

\section*{Acknowledgments}
This work is supported by Grant-in-Aid for Scientific research from the
Ministry of Education, Science, Sports, and Culture (MEXT), Japan, No.\ 22244021 (T.T.Y.),
No.\ 24740151 (M.I.), and also by the World Premier International Research Center Initiative (WPI Initiative), MEXT, Japan.
 The work of K.H.\ is supported in part by a JSPS Research Fellowship for Young Scientists.

\appendix

\section{Dynamical generation of mass scales by quark condensation}
\label{sec:IYIT}
In this appendix, we discuss the generation of the mass scales in Eq.~(\ref{eq:superpotential}) by quark
condensation in some hidden gauge sector. Let us consider
an $Sp(N_c)$ gauge theory with massless $N_c+1$ pairs of quark $Q_i$
($i=1,2,...,2(N_c+1)$) in the fundamental representation. Since this theory is
asymptotically free, we expect that the theory is well described by
gauge-invariant degrees of freedom at low energy. Actually, it is known
that the theory is described by $(N_c+1)(2N_c+1)$ mesons ${\cal M}_{ij}\sim
Q_iQ_j$ fulfilling the quantum constraint
\begin{eqnarray}
 {\rm Pf} \,{\cal M}=
2^{N_c-1}\Lambda^{2(N_c+1)},
\label{eq:}
\end{eqnarray}
where $\Lambda$ is the dynamical scale of this
$Sp(N_c)$ gauge theory \cite{Seiberg:1994bz}. Therefore, a quark condensation, $\vev{QQ}\sim
\vev{{\cal M}}\sim \Lambda^2$, occurs in this setup. This condensation can
provide us with the
mass parameters in Eq.~(\ref{eq:superpotential}).
For that purpose, we suppose that there are two independent hidden gauge
sectors $Sp(N_c)$ and $Sp(N_c')$ which generate $\mu^2 = k_1 \langle\cal{O}\rangle$
and $m$, $m'$, respectively. Let us discuss the
generation of $\mu^2$ and $m$, $m'$ in turn.

\subsection{Generation of $\mu^2$}
First note that the constraint does not fix the field values of the mesons
and the mesons hence remain massless.
In order to fix the field values, we introduce
$(N_c+1)(2N_c+1)-1$ $Sp(N_c)$-singlet chiral multiplets $T_{p}$ and assume a superpotential
of the following form
\begin{eqnarray}
W=c_{p,ij}T_pQ_iQ_j \sim c_{p,ij}T_p {\cal M}_{ij},
\label{eq:superpotential-app1-1}
\end{eqnarray}
where $c_{p,ij}$ are coupling constants.
Here, we have suppressed the gauge indices. 
There are $(N_c+1)(2N_c+1)-1$ $F$-term conditions for the $T_p$ fields and one
quantum constraint, while there are $(N_c+1)(2N_c+1)$ meson fields.
Therefore, for generic coupling constants $c_{p,ij}$, this
superpotential completely fixes the
meson fields in the quantum moduli space.

The quark condensate breaking the $R$ symmetry corresponds to the
expectation value of the operator
\begin{eqnarray}
 {\cal O}= \frac{1}{\mpl^{2n-3}} (QQ)^n
\label{eq:superpotential-app1-2}
\end{eqnarray}
in the superpotential.
Here, summations over gauge and flavor indices are implicitly understood.
This operator ${\cal O}$ actually condenses and corresponds to the operator ${\cal O}$ in
Eq.~(\ref{eq:cosmological constant}).
The genericity of the superpotential in Eqs.~(\ref{eq:superpotential-app1-1}) and (\ref{eq:superpotential-app1-2})
is guaranteed by the $R$ charges of $Q[1/n]$ and $T[2-2/n]$ with $n>2$.%
\footnote{Note that these charge assignments imply a gauge anomaly, which is
a necessary condition to achieve $R$ symmetry breaking without breaking SUSY~\cite{Dine:2009sw},
see also footnote~\ref{fn:discreteR}.}
As we have discussed in section~\ref{sec:model}, the Polonyi field $S'$ is
coupled to the quark condensate in the superpotential $W=k_1/\mpl S'\langle{\cal O}\rangle$ such that
$\mu^2$ can be identified as
$\mu^2\sim \Lambda^{2n}/\mpl^{2n-2}$ with $\Lambda$ being the dynamical scale of the gauge theory.

In this letter, we couple the Polonyi field in addition to the pseudo moduli $X$ and
$\bar{X}$.
The full superpotential for the fields $S$, $X$ and $\bar{X}$ is then given by
\begin{eqnarray}
 W= \mu^2S' + k_2 (S'-M)X{\bar X} + mX{\bar Y} + m'{\bar X} Y.
\end{eqnarray}
The Polonyi field $S$ in the main text and $S'$ are related by $S'=S+M$.
In order to solve the Polonyi problem, we utilize the vacuum with
$\vev{S'}=M~(\vev{S}=0)$ during and after inflation so that $X$
has a small mass and rolls down to a large field value. Even if $S'$ has
such a large field value, the dynamics of the hidden quarks are not affected
since the coupling between the Polonyi field and quarks is
suppressed by the Planck scale.

\subsection{Generation of $m$ and $m'$}
The terms $mX\bar{Y}$ and $m'\bar{X}Y$ are generated
in a similar way as the $\mu^2$ term discussed in the previous subsection.
We introduce $(N'_c+1)(2N'_c+1)-1$ $Sp(N'_c)$-singlet fields
$T'_p~(p=1,2,...,(N'_c+1)(2N'_c+1)-1)$
and assume the superpotential
\begin{eqnarray}
 W = c'_{p,ij}T'_p Q'_iQ'_j
+\frac{y_{ij}}{\mpl}X\bar{Y} Q'_iQ'_j + \frac{y'_{ij}}{\mpl}\bar{X}Y Q'_iQ'_j.
\end{eqnarray}
For the $Z_M$ symmetry to be broken, we assume that the $Q'$ quarks have $Z_M$
charges~$-1/2$.
The mass parameters $m^{(')}$
are then given by $m^{(')}=y^{(')} \Lambda'^2/\mpl$, where $y^{(')}$ are
appropriate combinations of the $y^{(')}_{ij}$ and where $\Lambda'$ is the
dynamical scale of the gauge theory.

In order to solve the Polonyi problem, we assume that $X$ has a large
field value during inflation. In this case, $\bar{Y}$ and $T'_p$ combine
to work as a SUSY breaking field in the model discussed in Ref.~\cite{Izawa:1996pk}.
Therefore, there is an extra contribution  to the vacuum energy
during inflation, $\Delta V \sim \Lambda'^4 \sim m^2 \mpl^2$.
Since we assume that $m<H_{\rm inf}$, this vacuum energy does not affect
the inflationary dynamics.

\end{document}